\documentclass[conference]{IEEEtran}
\IEEEoverridecommandlockouts
\usepackage{amsmath}
\usepackage[caption=false]{subfig}
\usepackage{tabularx,graphicx}
\usepackage{hyperref}
\usepackage[numbers,sort&compress]{natbib}
\usepackage{adjustbox}
\usepackage{tabularx,multirow}
\usepackage{xcolor}

\begin{document}

\title{Player Modeling using Behavioral Signals in Competitive Online Games}


\author{\IEEEauthorblockN{Arman Dehpanah}
\IEEEauthorblockA{\textit{School of Computing} \\
\textit{DePaul University}\\
Chicago, USA \\
\small{adehpana@depaul.edu}}
\and
\IEEEauthorblockN{Muheeb Faizan Ghori}
\IEEEauthorblockA{\textit{School of Computing} \\
\textit{DePaul University}\\
Chicago, USA \\
\small{mghori2@depaul.edu}}
\and
\IEEEauthorblockN{Jonathan Gemmell}
\IEEEauthorblockA{\textit{School of Computing} \\
\textit{DePaul University}\\
Chicago, USA \\
\small{jgemmell@cdm.depaul.edu}}
\and
\IEEEauthorblockN{Bamshad Mobasher}
\IEEEauthorblockA{\textit{School of Computing} \\
\textit{DePaul University}\\
Chicago, USA \\
\small{mobasher@cs.depaul.edu}}
}
 
\maketitle

\begin{abstract}

Competitive online games use rating systems to match players with similar skills to ensure a satisfying experience for players. In this paper, we focus on the importance of addressing different aspects of playing behavior when modeling players for creating match-ups. To this end, we engineer several behavioral features from a dataset of over 75,000 battle royale matches and create player models based on the retrieved features. We then use the created models to predict ranks for different groups of players in the data. The predicted ranks are compared to those of three popular rating systems. Our results show the superiority of simple behavioral models over mainstream rating systems. Some behavioral features provided accurate predictions for all groups of players while others proved useful for certain groups of players. The results of this study highlight the necessity of considering different aspects of the player’s behavior such as goals, strategy, and expertise when making assignments.

\end{abstract}


\section{Introduction}
\label{sec:introduction}

Online gaming has become one of the most popular entertainment platforms over the past decade.
Games such as Apex Legends, PlayerUnknown's Battlegrounds, and League of Legends serve millions of players on a daily basis.
These games consist of groups of players competing against each other or working together to accomplish a task.

Many online games use an extension of the famous Elo system~\cite{elo1978rating} for assigning players to teams and creating match-ups.
Traditionally, these systems solely focus on players' skill level and let players with the same level of skills play against each other.
The skill level of players is represented by a skill rating.
Such ratings are used to test match-ups and create leaderboards.
While simple and fast, using skill as the only important factor for team assignments does not guarantee a positive experience for players~\cite{delalleau2012beyond}.
Other systems attempt leveraging different in-game statistics to predict the outcome of potential match-ups and create balanced teams accordingly.
These systems often require high computation power due to their memory-intensive and complex computations~\cite{bakkes2012player}.
In addition, compared to traditional rating systems, these systems often lack explainability and are difficult to understand for regular players~\cite{jacob2020s, seif2020data}.

There has been an increasing amount of research on extending such approaches.
However, many introduced systems are either too specific, meaning they can only be applied to a specific game or genre, or are too global, meaning they do not exploit many relevant data available through different sources such as in-game statistics or social network of players.
Players differ in terms of behavioral aspects such as expertise, strategy, intentions, and goals.
It is particularly important to address these aspects in team-based games since such differences may influence the interactions between players and as a consequence sway the team performance.
In some cases, it is better to match players who share the same features.
For example, players interested in exploration and discovery should be grouped together, and not grouped with players interested in high-level competition.
In other cases, players should be grouped based on their differences.
For example, each team should have a player willing to take on the leadership role, but we do not want a team of five players vying for leadership.

This paper is the first step in our research agenda to extend upon past efforts by introducing a comprehensive framework for modeling players and making assignments.
In this paper, we seek to answer the following questions: \:\emph{(1) How to model players?\: (2) What defines a reliable player model?\: (3) How to use player models for evaluating match-ups?}

To answer the first question, we perform extensive feature engineering using in-game statistics.
We then explain how these features capture different behavioral aspects of players to answer the second question.
Online games have a very clear evaluation measure, the ranks of players or teams in a match.
To answer the third question, we use created player models to predict ranks and evaluate match-ups.

We limit our experimentation to battle royale games; a popular game mode that has not received enough attention in the research literature despite its increasing popularity.
We are concerned with how to model a single player, a necessary step before modeling a team.
Therefore, we focus on solo battle royale where many players compete against one another as singletons at the same time in the same match, aiming to be the last one standing to win the game. 

We hypothesize that players can be described by their playing behavior, and that we can achieve more accurate rank predictions when building our prediction models based on behavioral signals.
We compare our prediction results with those of three popular rating systems in three different setups, each of which corresponds to a certain aspect of the game-play.
Our results imply the high potential of simple behavioral features for modeling players and predicting their ranks.
The predictions obtained from these features are, in many cases, more accurate than those of state-of-the-art algorithms.
The results of this study will be used to extend upon previous works by introducing a generalizable player modeling framework that covers relevant aspects of players' behavior.

The rest of this paper is organized as follows:
In Section~\ref{sec:relatedWork}, related algorithms and methods are explained.
In Section~\ref{sec:method}, the methodology is explained in detail.
In Section~\ref{sec:preliminary}, the preliminary results are discussed.
Finally, we conclude the paper in Section~\ref{sec:plans} by mentioning the future directions of our research agenda.

\section{Related work}
\label{sec:relatedWork}

Online games are quickly surpassing offline games in the number of players~\cite{website1}.
These games are particularly popular because of the embedded social interactions in co-up tasks and competitions.
To create an engaging and satisfying experience for players, these games use matchmaking systems to assign players to matches.

Many matchmaking systems work based on generating balanced matches.
The most prominent examples of such systems are Elo~\cite{elo1978rating}, Glicko~\cite{glickman1995glicko} and TrueSkill~\cite{herbrich2007trueskill}.
These systems leverage players' previous match results to assign them a skill rating and then, let the players with similar ratings play against each other.
The ratings are updated after each match based on the match result.
Such systems are referred to as rating systems.
Besides matching players, the ratings are also used for creating leaderboards and providing players with a tool for following their skill development.
Despite their widespread use, these systems, in their traditional form, suffer from a provincial outlook, i.e., rating players solely based on the outcome of the match.

Several works extended upon traditional rating systems by considering different aspects of game-play such as context~\cite{zhang2010factor, chen2016predicting}, chemistry between team members~\cite{delong2011teamskill}, skill aggregation~\cite{dehpanah2021evaluating}, offensive/defensive skills~\cite{guo2012score, stanescu2011rating}, victory margin~\cite{kovalchik2020extension}, draw margin~\cite{dangauthier2007trueskill}, experience~\cite{minka2018trueskill}, and time~\cite{coulom2008whole}.
Efforts have also been made to introduce new matchmaking algorithms focusing on aspects such as competitive balance ~\cite{nikolakaki2020competitive, delalleau2012beyond}, playing style~\cite{wang2015thinking}, and player engagement and retention~\cite{stroh2018more, chen2017eomm}.
The majority of such algorithms use neural networks and latent factor models to predict rank and create match-ups.
While these models often achieve better prediction results and higher quality matchmakings compared to rating systems, they lack explainability due to the complex nature of their calculations; an important feature for players who are interested in understanding the logic behind their match-ups~\cite{jacob2020s, seif2020data}.

Creating balanced match-ups is the main objective of many matchmaking algorithms.
Although having balanced matches seems crucial in competitive online games, it has been argued that focusing solely on players' skills and balance does not guarantee a positive gaming experience for all players~\cite{nikolakaki2020competitive, delalleau2012beyond}.
Instead, such a positive experience is directly dependent on players' behavior and preferences~\cite{riegelsberger2007personality}.

In this paper, we focus on achieving a better understanding of players' behavior.
To this end, we conduct extensive experimentation on a large real-world dataset to engineer several features that cover different aspects of players' behavior.
We then use those features to predict their ranks.

Our work differs from previous efforts.
We perform our experimentation on battle royale matches, while the majority of past efforts focus on head-to-head matches.
We engineer several features from players' behavioral signals covering different aspects of their game-play.
While previous works often model players using heavy mathematical computations and statistical estimations, we model players using engineered behavioral features and create simple predictive models based on each individual feature to predict ranks.
We compare the prediction results with those of three popular rating systems, i.e., Elo, Glicko, and TrueSkill.


\section{Methodology}
\label{sec:method}

In this section, we introduce the dataset used to perform our experiments along with the engineered behavioral features.
We then detail our experimental setup and explain how we evaluate rank predictions.

\subsection{Dataset}
PlayerUnknown's Battlegrounds (PUBG) is a popular shooter battle royale video game pitting up to one hundred players against each other to scavenge for weapons and fight for eliminating opponents.
The game includes a safe zone that decreases in size over time to force players to engage in fights.
The goal is to survive until the very end and be the last player standing to win the match.
PUBG is played in teams of four (squad), teams of two (duo), and solo where players compete against each other as singletons.
The dataset is publicly available on \textit{Kaggle}\footnote{https://www.kaggle.com/skihikingkevin/pubg-match-deaths}.
Since this study focuses on modeling single players, we only considered solo matches in the dataset.
The filtered dataset provides in-game statistics such as the number of kills, rank, and distance walked for over 75,000 matches and 1,700,000 unique players.

\subsection{Behavioral Features}
In this section, we introduce a selection of behavioral features extracted from in-game statistics and explain how we model players using those features.
Each extracted feature corresponds to a certain aspect of the playing behavior of the player.
All features are calculated at the start of each match based on historic data, and are updated after each match based on player activities within that match.
From raw statistics such as player's name, rank, number of kills, damage dealt, distance walked, and time survived in each match, we engineered the following features:

\textbf{$\beta_1$: number of games played.} Every game has a learning process.
We assume that the more players play a game, the better they become in that game by gaining more experience. Upon release of a game in a given genre, players who already played several games from that genre might have an advantage over the players who are new to that genre.
However, even games from the same genre could be remarkably different due to particular game-play features such as characters with unique abilities or weapons with unique attributes.
Therefore, our assumption would still hold.
The number of games played is a good indicator of how much effort players put into the game and how engaged they are with the game.
    
\textbf{$\beta_2$: kill to death ratio (K/D).} The number of kills in a player's profile is not always indicative of their playing style or skills.
For example, a player may have 100 kills in their profile but they may have scored those kills over hundreds of games.
Leveraging kill to death ratio alleviates this issue by considering the total number of kills players scored in relation to the total number of times they were defeated in their matches.
Players with higher K/D are assumed to be more skilled and aggressive in their playing style.
    
\textbf{$\beta_3$: firing accuracy.}
Experienced and skilled players are assumed to have higher firing accuracy as they know how and where to shoot to get the best results.
This measure can be calculated as the ratio of the total number of kills to the cumulative amount of damage dealt over the games played.
    
\textbf{$\beta_4$: survive ratio.} The main goal of battle royale matches is to survive or be the last player standing.
While aggressive and high-skilled players may achieve this goal by attempting direct fights to eliminate opponents, others may adopt a more passive strategy and avoid fights until forced by the shrinkage of the map's safe zone, and thus, survive until the very end.
For each player, we calculate survive ratio as the cumulative amount of time survived in their matches divided by the total number of matches they played.
This feature may represent both expertise and strategy, especially if considered together with features such as K/D or firing accuracy.
    
\textbf{$\beta_5$: walking ratio} and \textbf{$\beta_6$: riding ratio.} The distance traveled during a match is often indicative of a player's strategy.
For example, aggressive or high-skilled players are often in motion for the whole duration of the match trying to find and fight opponents to achieve more kills.
On the other hand, `campers' often prefer to choose a safe spot to hide and wait for ambushing other players.
While walking distance directly shows such strategies, riding distance may imply a different story.
In-game vehicles can be used for several reasons.
For example, upon the safe area shrinkage, experienced players may stay calm and escape the unsafe zone faster by using a vehicle while amateur players are often baffled quite easily upon receiving damage outside the safe zone and rush towards the safe zone on foot.
Also, aggressive players often use vehicles to reach unlooted areas faster, find potential opponents, or attempt a `third-party', i.e., attacking opponents who already weakened each other in their ongoing fight.
Finally, players interested in discovery often use vehicles for fun to explore the map and enjoy the non-competition aspect of the game.
We calculate these features as the ratio of cumulative walking and riding distance traveled to the total number of games played.
    
\textbf{$\beta_7$: walking velocity} and \textbf{$\beta_8$: riding velocity.} Since battle royale games often include many players, it is very common that several players are eliminated very early in the match even before finding a weapon.
In these cases, features such as walking or riding ratio could be misleading.
To alleviate this issue, we consider the cumulative distance traveled by players in relation to the cumulative time they survived in the matches they played to calculate their walking and riding velocity.
These features are indicative of a player's strategy by showing how much time they spend on their matches traveling on foot or by vehicles.
    
\textbf{$\beta_9$: rank ratio.} A rank is not necessarily indicative of a player's overall expertise.
Players may show different performance levels under different circumstances.
For example, average players may seem like skilled players in a match where most of their opponents are new players who started playing that game recently.
Since weapons and equipment are randomly scattered over the map, even highly skilled players may be eliminated early in the game if they cannot find a suitable weapon and are faced with a group of amateur opponents with powerful weapons.
Since each match involves different number of players, we calculate rank percentages for each player in the match.
We then divide their cumulative rank percentages by the total number of games they played to calculate their rank ratio.
    
We create simple player models based on each feature and use those models to predict rank.
In a field \textit{F}, denoting a battle royale match where \textit{n} players $p_1$, $p_2$, ..., $p_n$ compete against each other, the prediction function is defined as:

\begin{equation*}
    \Phi =  \operatorname*{arg\,sort}_{i \in \{1,2,...,n\}} \Big((F,p_i)\: |\: \beta\Big)
\end{equation*}  

\noindent where $\beta$ is the behavioral feature of interest.
For all features, we assume that higher values correspond to better ranks, except for rank ratio where smaller values correlate to better ranks.
The prediction function takes the list of the players competing in a battle royale match along with their corresponding behavioral feature and sorts players based on their feature value in a descending order.
The ties are randomly broken by the model.
The resulted order is then returned by the function as predicted ranks for that match.
The predicted ranks are then compared with the observed ranks for each match to evaluate the performance of the models.



\begin{table*}[htbp]
  \centering
  \caption{The average predictive performance of rating systems and behavioral models. \:$\beta_1$: number of games played, \:$\beta_2$: K/D,  \:$\beta_3$: firing accuracy, \:$\beta_4$: survive ratio, \:$\beta_5$: walking ratio, \:$\beta_6$: riding ratio, \:$\beta_7$: walking velocity, \:$\beta_8$: riding velocity, and \:$\beta_9$: rank ratio.}
\begin{adjustbox}{width=500 pt,center}  \begin{tabular}{l c c c c c c c c c c c c}
    \hline
    \hline
    Setup (\%NDCG) & Elo & Glicko & TrueSkill & $\Phi(\beta_1)$ & $\Phi(\beta_2)$ & $\Phi(\beta_3)$ & $\Phi(\beta_4)$ & $\Phi(\beta_5)$ & $\Phi(\beta_6)$ & $\Phi(\beta_7)$ & $\Phi(\beta_8)$ & $\Phi(\beta_9)$\\ 
    \hline
    
     All Players & 56.8 & 56.2 & \textcolor{red}{57.4} & 54.1 & \textcolor{blue}{60.1} & 56.1 & \textcolor{blue}{59.7} & \textcolor{blue}{58.6} & \textcolor{blue}{58.5} & 55.9 & 56.8 & \textcolor{blue}{\textbf{61.3}}\\
     
     Top-tier Players & 73.7 & 62.4 & \textcolor{red}{79.1} & 59.0 & 71.6 & 57.8 & \textcolor{blue}{79.4} & 69.6 & 67.8 & 56.7 & 57.3 & \textcolor{blue}{\textbf{85.1}}\\
     
     Frequent Players & 59.3 & \textcolor{red}{63.8} & 57.9 & \textcolor{blue}{\textbf{86.4}} & 60.7 & 57.6 & 59.9 & 58.9 & \textcolor{blue}{64.1} & 58.1 & \textcolor{blue}{64.2} & 63.0\\
    \hline
    \hline

  \end{tabular}
  \end{adjustbox}

  \label{tab1}
\end{table*}


\subsection{Experimental Setup}

In this paper, we compared the prediction results of our simple behavioral player models with those of three popular rating systems; Elo, Glicko, and TrueSkill.
We used TrueSkill in its original form with the parameters recommended in its official documentation while using the extended version of Elo and Glicko for battle royale games~\cite{arman2021evaluation}.
We sorted the matches by their timestamps. For each match, we retrieved the list of players.
New players – those that have yet to appear in a match – were assigned default
values; 1500 for Elo and Glicko, 25 for TrueSkill, 100 for rank ratio, and zero for all other behavioral features.
We then sorted the players based on these values and considered the created ordering as predicted ranks.
Predictions were then compared with observed ranks. 

The outcome of all tested models is an ordered list of players showing their predicted ranks.
Therefore, we used an extension of Normalized Discounted Cumulative Gain (NDCG) for evaluating rank predictions~\cite{dehpanah2021evaluation}.
We performed our experiments in three different setups.

In the first setup, we considered all players in the dataset regardless of how skilled they are or how many games they played.
This setup includes players who are new to the system appearing for the first time in the data.
These players are assumed to be new players.
This setup also includes players who only played a few matches.
These players are assumed to be seasonal players who are not engaged with the game.
Rank predictions are often hampered in presence of such players.

In the second setup, we considered top-tier players in the dataset.
To identify these players, we selected 500 players with the highest win rate (ratio of the number of games won to the total number of games played) who also played more than 10 games.
We evaluated rank predictions on their first 10 games played.

The last setup includes the most frequent players in the game.
To identify these players, we selected all players who played more than 100 games.
We evaluated rank predictions on their first 100 games played.


\section{Results and discussions}
\label{sec:preliminary}

\begin{figure*}
   \centering
\includegraphics[width= 18cm]{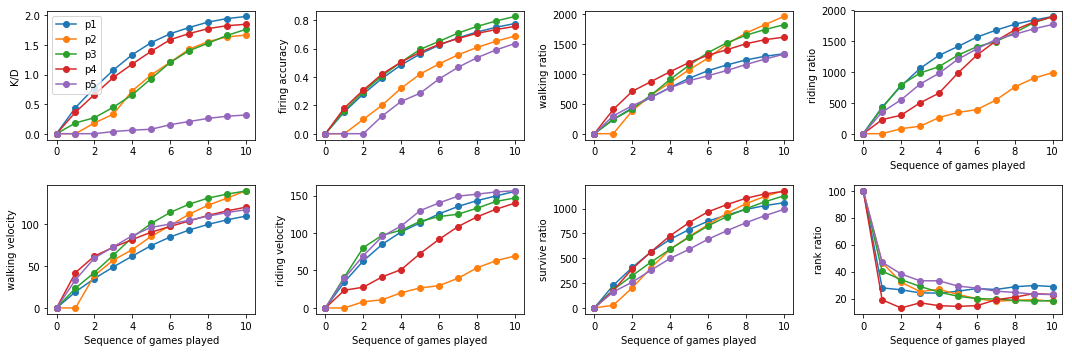}
\caption{The development of behavioral features for five players in the dataset with the highest win rate.}
\label{fig:top} 
\end{figure*}

\begin{figure*}
   \centering
\includegraphics[width= 18cm]{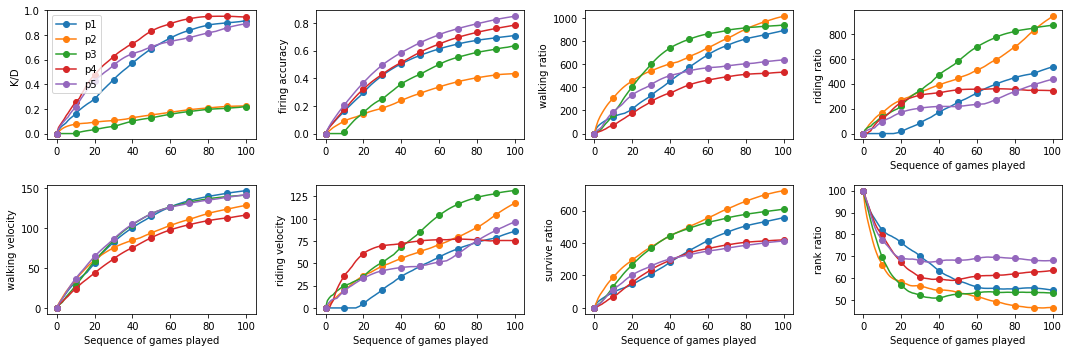}
\caption{The development of behavioral features for five players in the dataset with the highest number of games played.}
\label{fig:freq} 
\end{figure*}

In this section, we discuss the rank prediction results.
The results are shown in table~\ref{tab1} as well as figures~\ref{fig:top} and~\ref{fig:freq}.
Table~\ref{tab1} shows the average NDCG achieved by behavioral models and rating systems for each experimental setup.
In each setup, the best NDCG value for rating systems is shown in red color while blue-colored values show the cases where behavioral models outperformed rating systems.
Also, the highest NDCG value for each setup is shown in bold.

Based on the results of the `all players' setup, the highest NDCG value belongs to rank ratio.
Interestingly, several other behavioral models including K/D, survive ratio, walking ratio, and riding ratio also achieved more accurate predictions than rating systems in this setup.
As mentioned before, this setup involves new and seasonal players who are not known to the system.
Rating systems need to observe several games from players to achieve a reliable estimation of their performance.
However, based on the results, simple behavioral features seem to better and faster capture the true performance level of players assisting them to achieve more accurate predictions for this scenario. 

For the `top-tier players' setup, again, the best performance belongs to rank ratio with an excellent NDCG of 85.1\%.
While TrueSkill achieved a high NDCG value of 79.1\% outperforming Elo and Glicko by a fairly large margin, it was surpassed by another behavioral model, i.e., survive ratio with 79.4\% NDCG.
Rating systems learn the true skill level of top-tier players faster compared to other players because these players often show consistent playing behavior and decrease the system's uncertainty about their skills.
As shown in table~\ref{tab1}, the predictions generated by these systems are more accurate compared to their predictions for the `all players' setup.
However, based on the results, simple behavioral models such as rank ratio and survive ratio outperformed rating systems in this task.

For the `frequent players' setup, Glicko achieved the best results among the three mainstream rating systems.
However, it was outperformed by the number of games played with a large margin.
This behavioral feature achieved 86.4\% NDCG, the highest value among all models in all setups.
Riding ratio and riding velocity also outperformed rating systems in this scenario.
Rating systems are popular for learning more about players and achieving better predictions as they observe more games from them.
Comparing the NDCG values for this setup with those of the `all players' setup confirms such a case.
However, simple behavioral models such as the number of games played achieved much higher NDCG values, showing their capability of doing such a task more accurately.


To show the ability of behavioral features in capturing the true performance of players who show consistent playing behavior and experienced players who are more engaged with the game, we plot the development of behavioral features over time for top-tier and frequent players.
Figures~\ref{fig:top} and~\ref{fig:freq} show time series of development patterns for the top-tier and frequent players, respectively.
For legibility, this analysis was done on a small sample of players representing each group.
For top-tier players, we selected the top five players having the highest win rate and showed their development over the course of their first 10 games.
For frequent players, we selected the top five players with the highest number of games played and demonstrated their development over the course of their first 100 games.

As can be seen in the figures, the behavioral features for all of these players develop consistently over time at different rates.
For example, some top-tier players obtain K/D of 0.3 after the 10\textsuperscript{th} game while others achieve K/D of 2.0 at the end of the sequence.
Also, some frequent players show high riding and walking velocity in their games while others choose more passive strategies.
Based on the rank ratio values, most of the top-tier players rank below the 30\textsuperscript{th} percentile after their second game while most of the frequent players often remain in the 50\textsuperscript{th} rank percentile.
The observed patterns certainly imply the ability of the selected behavioral features in modeling the performance of players.

\section{Conclusion and Future work}
\label{sec:plans}

In this paper, we proposed several features engineered from in-game statistics for modeling the behavior of players and used those behavioral models for predicting their ranks.
Predicted ranks were then compared to those of three popular rating systems; systems that model players mainly based on their skill level with respect to their achieved ranks in the matches they played.
Our results showed that, despite their simplicity, behavioral models achieve more accurate predictions than mainstream rating systems.
The behavioral features also accurately demonstrated the development of players over time implying the capability of such features in capturing the true performance level of players.

The results of this study highlighted the necessity of scrutinizing how players are modeled and how player models are used to predict their ranks and create match-ups.
In the next phase of our research agenda, we plan to create predictive models based on a weighted hybrid of the engineered behavioral features.
We also plan to combine the strengths of behavioral models and rating systems to provide explainable ratings and match-ups based on the players' behavior.
To that end, we will embed behavioral features in a weighted hybrid and use them as inputs for the rating systems.
Such models are scalable so long as each of the components is scalable, a relatively easy task when each component is responsible for only a single aspect of the player's behavior.
In addition, such models are easily extensible; if a new feature is proposed, it can simply be appended to the models.



\bibliography{main}
\bibliographystyle{unsrt}

\end{document}